\documentclass{rspublic}

\usepackage{graphicx} 
\usepackage[hang,center]{subfigure}
\begin{document}

\title[Multifractal analysis of a road-to-crisis in a Faraday experiment]{Multifractal analysis of a road-to-crisis in a Faraday experiment}

\author[M. Rosen, M. Piacquadio]{M. Rosen, M. Piacquadio}

\affiliation{Laboratorio de Medios Porosos, Facultad de Ingenieria, Universidad de Buenos Aires, Paseo Col\'{o}n 850, (1063) Buenos Aires, Argentina}

\label{firstpage}

\maketitle

\begin{abstract}{Multifractality, Faraday experiment, Chaos}
A thin layer of liquid in a horizontal cell is subjected to a periodic vertical force with two control parameters: acceleration and frequency. The influence of the rheological behavior of the fluid was considered over the empirically obtained results, which were subjected to a multfractal spectrum process. As control parameters varied, so did the corresponding spectra, according to theoretical models; each change of state was visually detected, which permitted interpreting the corresponding spectral changes.
\newline
\newline
\end{abstract}
\section{Introduction}
The classical Faraday patterns are a selection of surface waves following a specific criterion studied by Benjamin and Ursell (1954). The amplitude and the wave number of the surface follow a certain harmonic oscillator equation. In this paper, the liquid is "shear thinning", adding further complexity to the classical case. A thin layer of liquid in a horizontal cell is subject to a periodic vertical force, its frequency and acceleration being the control parameters. 
Between the acceleration/frequency threshold (acceleration measured in $g$ units\textemdash $g$ being the acceleration due to gravity) after which the classical Faraday patterns appear, and the new threshold corresponding to the ejection of droplets (spray), there is an intermediate zone with a structure that is the object of our study. This intermediate state is characterised by the formation of \textquoteleft cusps\textquoteright \  on the surface of the liquid under study; the behaviour of the geometrical patterns formed by these \textquoteleft cusps\textquoteright \  depends on the control parameters of the system.
The liquid used in these experiments is a polymeric solution of polyacrylamide (PAAm) in a $2$g/l concentration. PAAm is a shear-thinning polymer, i.e. there is a region where its viscosity diminishes with the shear rate. 
In this paper, the behaviour of the \textquoteleft cusps\textquoteright \  is studied by imposing a voltage difference across a pair of electrodes. One electrode is submerged in the liquid; the other is suspended over the liquid surface at a controllable distance. 
When a cusp rises and touches the electrode above, it closes a circuit and a current is generated yielding a signal, electronically detected and processed. In figure 1 we present the experimental setup and figure 2 shows the images corresponding to three different frequencies and the response of the detector, where we can observe the signal as a function of time.
With constant frequency $\nu$, we vary the acceleration $\kappa$ and obtain a signal as a function of time. We study $C_\kappa$, the Cantordust intersection of the signal with a segment.
\newline\indent
In the literature there are two ways to obtain a multifractal spectrum (alpha, f(alpha)) of a Cantordust: one is through the very definitions of the alpha-concentration and the corresponding spectral value f(alpha), the other is through the use of the lagrangian coordinates. It is not proved that the two methods yield the same spectrum. The spectrum given by the lagrangian coordinates is an ideal one, to which the spectrum given by the definitions is supposed to approximate. Yet, the latter can have deviations from the ideal spectrum. In this paper, we work with the definitions (despite the difficulties involved, e.g. far less points in the spectral curve), since such deviations from the ideal shape give useful information on the nature of the experiment.
\section{Experimental setup}
The experimental setup consists of a square acrylic container (side 7 cm), on top of a vibrator. The system is excited in the vertical direction with an acceleration $\kappa = a_0 \sin \nu t$, where $\nu$ is the driving frequency and $a_0$ the other control parameter. A pair of electrodes, one submerged at the bottom of the cell, the other 3 mm from the surface, permits detecting the contact of the liquid with the second electrode, thus closing an electric circuit and sending a time signal.
\newline
\indent
The frequency applied varies between 20 and 120 Hz, the thickness of the liquid layer being 4 mm. 
\newline
\indent
As noted in the Introduction, the polymeric solution is a shear thinning liquid. We can observe (Fig. 3) that the viscosity diminishes with the shear rate. In Fig. 2 we observe the formation of cusps, and intermediate stage between pattern formation and droplet ejection.
\begin{figure}
\begin{center}
\subfigure{
    \includegraphics[clip,height=2.05in,width=5.0in]{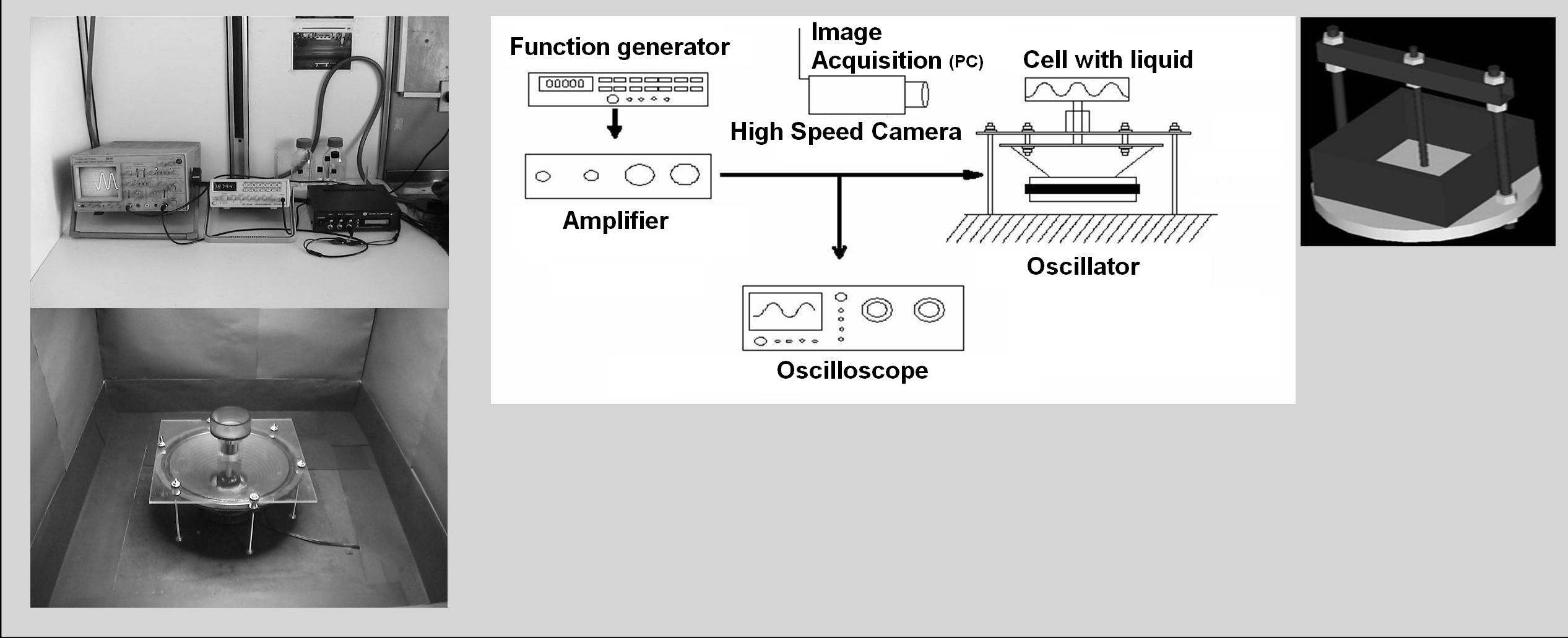} 
	  \label{fig1a}
}
\caption{Experimental setup}
\end{center}
\end{figure} 
\newline
\begin{figure}
\begin{center}
    \includegraphics[clip,height=2.05in,width=5.0in]{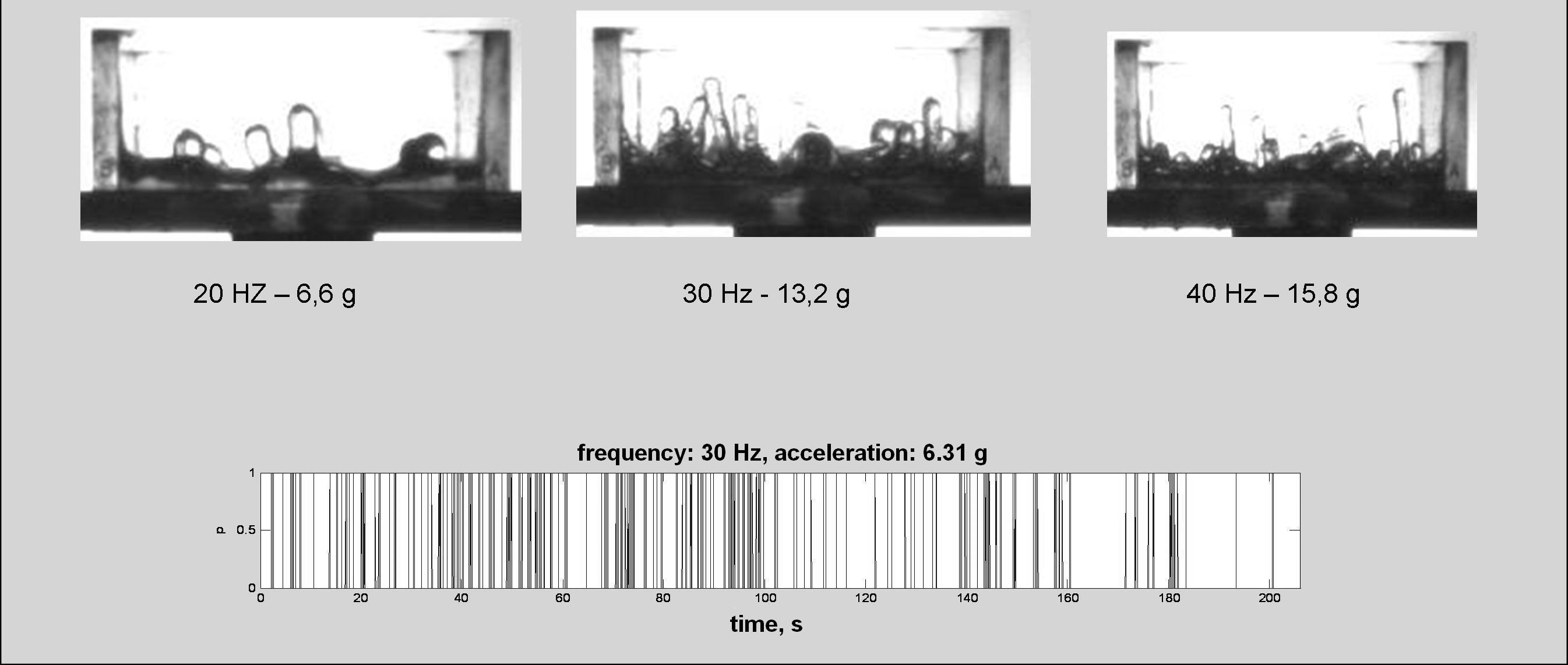} 
	  \label{fig1b}
\caption{Cusp images and obtained signal}
\end{center}
\end{figure}
\begin{figure}
\begin{center}
    \includegraphics[clip,height=2.05in,width=5.0in]{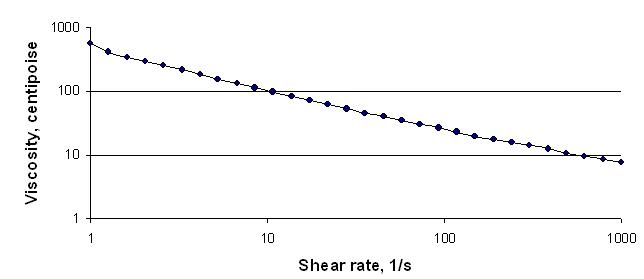} 
	  \label{h}
\caption{ Viscosity (centipoise) as a function of shear rate (1/s) for the experimental liquid PAAm, 2g/l}
\end{center}
\end{figure}
With constant frequency $\nu$, we vary the acceleration $\kappa$ and obtain a time signal. $C_\kappa$ is the Cantordust intersection of the signal with a segment. 
\section{Multifractality in an empirical context}
We will work in empirical fractal sets $F$ contained in the unit segment $[0,1]$, endowed with a probability measure $\mu$. Let us recall that the $\alpha$-concentration of a box $B \subset [0,1]$ of length $l$ is the log/log version of the density \textemdash mass or measure per unit length\textemdash\space that is $\alpha(B) = \frac{\log\mu(F \cap B)}{\log l}$. The $\alpha$-concentration of a point in $F$ is defined (just like the density of a point) as the limit of $\alpha$'s of shrinking boxes containing said point. The spectrum $f(\alpha)$ is the box dimension of the elements in $F$ that share the $\alpha$-concentration.
\newline
\indent
When dealing with an experimental sample, an empirical Cantordust fractal set embedded in a unit segment, which we want to process in order to obtain a multifractal spectrum $(\alpha,f(\alpha))$, some practical difficulties appear. There are several rather severe discrepancies with the theory proper. On the one hand, a fractal Cantordust-like set has a non-numerable infinity of elements, whereas in any concrete empirical context, we have a finite number of measurements in the sample. Hence, the theoretical concept \textquotedblleft$\epsilon \rightarrow 0$" is rendered meaningless: say that $\mathbf{S}$ is the size of the sample taken at regular intervals, and recall that the Cantordust is embedded in a unit segment. Then $\frac{1}{\mathbf{S}} = \Delta s = \epsilon_s$, where \textquotedblleft s" stands for \textquotedblleft sample", is the smallest possible value of such $\epsilon$. Next, we cover the unit segment with boxes of a certain length $\Delta l$ in order to define the so-called natural probability  measure $\mu$. Theoretically, such natural measure is defined as the proportion of sample elements in a box of length $\Delta l = \epsilon_l$ \textquotedblleft as $\epsilon_l \rightarrow 0$"\dots again impossible in an empirical context.  In such a context we must have $\epsilon_s$ considerably smaller than $\epsilon_l$, i.e. $\epsilon_s \ll \epsilon_l$. Therefore we need to have $\mathbf{S} \gg \mathbf{B}$, the total number of boxes of equal size. 
\newline
\indent
There is a yet more delicate $\epsilon$ to take into account: having $\epsilon_s$ and $\epsilon_l$, $\epsilon_s \ll \epsilon_l$, and the natural measure $\mu$, we establish the values $\alpha$ for each box $B, \alpha = \frac{\log(\mu(B \cap F))}{\log \epsilon_l}$.  We obtain thus a number of $\alpha$-values, from a certain $\alpha_{min}$ to an $\alpha_{max}$. Theoretically, we have to consider $f(\alpha) = \lim_{\epsilon_l\rightarrow 0}\frac{\log N_\alpha}{\log\frac{1}{\epsilon_l}}$, $N_\alpha$ the number of elements in the infinitesimal segment $[\alpha,\alpha+\epsilon_\alpha)$, $\epsilon_\alpha$ arbitrarily small and tending to zero. It may well be that the (\emph{finite} number of) $\alpha$-values empirically obtained between $\alpha_{min}$ and $\alpha_{max}$, are all different\dots\space in which case, if $\Delta\alpha=\epsilon_\alpha$ is rather small, each interval $[\alpha,\alpha+\Delta\alpha)$ may have \emph{one} element (if any)\dots\space in which case $f(\alpha)\equiv 0$.
\newline
\indent
In practice, it should be $\epsilon_s \ll \epsilon_l \ll \epsilon_\alpha$, though in theory all these $\epsilon$ tend to zero independently of one another. This implies that $\mathbf{S} \gg \mathbf{B}$ which in turn should be $\gg \mathbf{A}$, the number of points $(\alpha_i,f(\alpha_i))$ obtained in order to form a \emph{smooth} curve $(\alpha,f(\alpha))$. We have found, elsewhere (Piacquadio and Cesaratto 2001) and in this work, that in this context \textquotedblleft $\gg$" means \textquotedblleft larger than the square of", e.g. $\mathbf{S}$ should be larger than the $4^{th}$ power of $\mathbf{A}$. For instance, when calculating the multifractal spectrum of the Farey tree (Piacquadio and Cesaratto 2001) we needed an $\mathbf{S}$ of the order of $10^6$, whose $4^{th}$ root should yield 31 points $(\alpha_i,f(\alpha_i))$. Yet, we could obtain only some $15$ such points constituting a \emph{smooth} curve $(\alpha,f(\alpha))$ that agreed with the theoretical spectrum. 
\newline
\indent
The empirical context of this work admits a maximum value of $10^4$ elements in $\mathbf{S}$, for beyond that magnitude the experiment changes its nature, as seen in the last section.
\newline
\indent
Given that $\mathbf{S}=10^4$, then $\sqrt{10^4}=10^2$ has to be the maximum number $\mathbf{B}$ of boxes \textemdash the absolute maximum, a slightly smaller number would be more adequate. Hence, a number slightly smaller than $\sqrt{\mathbf{B}} = \sqrt{10^2} = 10$ points $(\alpha_i,f(\alpha_i))$ will yield a smooth curve $(\alpha,f(\alpha))$ in each of the cases below.
\section{The evolution of the Cantordust set $C_\kappa$}
In order to analyse the evolution of $C_\kappa$ as $\kappa$ increases, we adapt the model of the Chaos diagram for the one-parameter Henon-like system (Grebogi \emph{et al.} 1987; Alligood and Sauer 1993; Robert \emph{et al.} 1999). In this diagram the orbits of a one-parameter \textemdash denoted here by $\kappa$\textemdash\space family of Henon-like maps are represented, by means of the global bifurcations they undergo, as $\kappa$ slowly increases. Chaos is reached for a certain specific value $\kappa_0$, which means that the system is characterised by the presence of a one-piece connected fractal attractor $F_\kappa$. As variable $\kappa$ slowly increases beyond $\kappa_0$, the system in chaos undergoes certain key changes. Grebogi \emph{et al.} (1987), Alligood and Sauer (1993) and Robert \emph{et al.} (1999) studied the dynamics of the change \textemdash the evolution of the so-called accessible rotation numbers for the one-piece fractal attractor $F_\kappa$. When the increasing variable $\kappa$ reaches a certain critical value $\kappa_c$, the chaotic system breaks down, and undergoes a so-called \emph{boundary crisis}, in which the attractor $F_\kappa$ suddenly disappears. Some work has been done (Hansen and Piacquadio Losada 2005) in order to relate the evolution of the dynamics of this chaos-on-road-to-crisis model to the geometrical changes in the one-piece $F_\kappa$, as $\kappa$ approaches $\kappa_c$, in which $F_\kappa$ disappears and the system breaks down. Following this lead, we analyse the evolution of $C_\kappa$ by means of that of the geometry of $f_\kappa = f_\kappa(\alpha)$, the multifractal spectrum of the natural measure $\mu_\kappa$ associated with $C_\kappa$. The guiding principle may be stated as follows: we study fractal $C_\kappa$ by means of its natural measure $\mu_\kappa$, $\mu_\kappa$ by means of its multifractal spectrum $f_\kappa$, and the evolution of the system by the geometrical evolution of the curve $(\alpha,f_\kappa(\alpha))$.
\newline
\indent
Briefly, details follow, the chaos-on-road-to-crisis type of situtation, as $\kappa$ nears a critical value $\kappa_c$, is characterised by a one-piece connected multifractal spectrum $f_\kappa$, which fulfills all the right thermodynamic laws. When $\kappa$ reaches the value $\kappa_c$ the system reaches a crisis point and breaks down, as the Lagrangian coordinates fail to make sense. For $\kappa \geq \kappa_c$ the measure $\mu_\kappa$ seems to disappear, for the corresponding multifractal algorithm produces a function made up of unconnected pieces of what looks like fragments of different spectra \textemdash so the thermodynamic laws cease to be valid. 
\section{Chaos on road to Crisis}
For $\kappa$ near $\kappa_c=4.93g$, e.g. for $\kappa=4.08g$, the natural measure $\mu_\kappa$ of $C_\kappa$ produces a one-piece connected multifractal spectrum, as seen in Fig. 4. 
\begin{figure}
\begin{center}
    \includegraphics[clip,height=2.0in,width=2.0in]{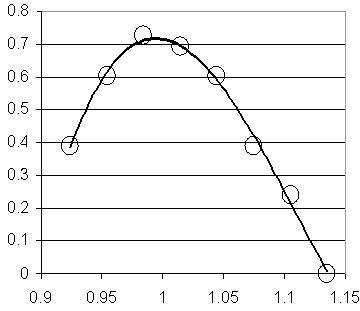} 
	  \label{fig2a}

\label{fig2}
\caption{The spectrum $f_\kappa(\alpha)$ for $\kappa < \kappa_c = 4.93g$ and frequency $\nu = 20$Hz}
\end{center}
\end{figure}
$C_\kappa$ was produced by a sample of size $\mathbf{S}=10^4;\mathbf{B}=\sqrt{\mathbf{S}}=100$, and $\mathbf{A}$ slightly smaller than $\sqrt{\mathbf{B}}$. We observe that the spectrum is convex, that there are points in the graph at both left and right of $\alpha=1$, that the maximal dimension is rather high \textemdash above $0.7$\textemdash\space in agreement with the naked-eye observation of the sample. The one law which does not appear to be fulfilled is $f(\alpha)$ being tangent to the line $y=x$, as shown in Fig. 5. This is due to the small value of $\mathbf{B}$. We increase the number $\mathbf{B}$ of boxes \textemdash setting at risk the smooth character of the graph of the spectrum\textemdash\space to 200. We observe that the convex character of the curve is preserved, as seen in Fig. 6, together with the features described above for $\mathbf{B}=100$. In table 1 we compare $\alpha_{min}$ for $\mathbf{B}=100$ and $\mathbf{B}=200$, we obtain $\alpha_{min}(200)=0.87\cdots <  \alpha_{min}(100)=0.90\cdots$. The same happens for the values of $\alpha_M$ for which the spectrum reaches its maximum: $\alpha_M(200)=0.94\cdots < \alpha_M(100)=0.96\cdots$. Both values indicate that, when $\mathbf{B}$ increases, the spectrum shifts towards the left, approaching the bisectrix line $y=x$. Also, $f_{max}(200)=0.74\cdots > f_{max}(100)=0.72\cdots$, which indicates that the shift is upwards as well as leftwards, getting sensibly closer to $y=x$. Increasing $\mathbf{B}$ further already distorts the smooth  character of the spectrum.
\begin{center}
\begin{figure}
\centering
    \includegraphics[clip,height=2.0in,width=2.0in]{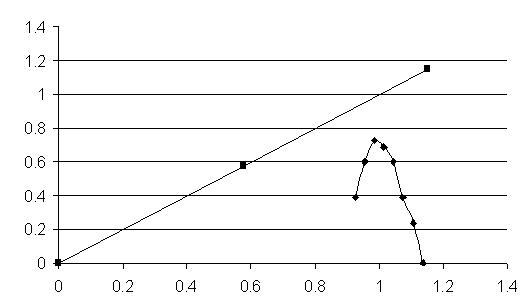} 
		\caption{The spectrum $f_\kappa(\alpha)$, $\kappa = 4.08g, 100$ boxes.}
    \includegraphics[clip,height=2.0in,width=2.0in]{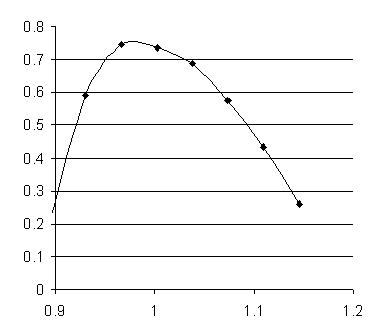} 
	\caption{The spectrum $f_\kappa(\alpha)$, $\kappa = 4.08g, 200$ boxes.}
\end{figure}
\end{center} 
\begin{center}
\begin{table}
  \begin{tabular*}{0.75\textwidth}%
     {@{\extracolsep{\fill}}cccr}
  --- & $\mathbf{B}=100$  & $\mathbf{B}=200$ \\
  \hline  
  $\alpha_{min}$ & 0.9091  & 0.8768   \\
  \hline  
  $\alpha_{M} / f(\alpha_M)$ is max.  & 0.9694  & 0.9485   \\
  \hline  
  $f_{max}$  & 0.7235  & 0.7457   \\
  \hline  
  $\alpha_{max}$ 1  & 1.120  & 1.1278   \\
  \end{tabular*} 
\caption{The spectrum $f_\kappa(\alpha)$, $\kappa = 4.08g,100$ and  $200$ boxes }
\end{table}
\end{center}
\section{Crisis: critical value $\kappa_c$ of $\kappa$}

The frequency $\nu=20$ Hz is kept constant. When $\kappa$ reaches the critical value $\kappa=\kappa_c=4.93g$, keeping $\mathbf{S}=10^4$ and $\mathbf{B}=\sqrt{\mathbf{S}}=10^2$, we obtain for $f(\alpha)=f_{\kappa_c}(\alpha)$ a graph with $f'(\alpha)$ constant along a large segment, as shown in Fig. 7. 
\begin{figure}
\begin{center}
\subfigure{
    \includegraphics[clip,height=2.0in,width=2.0in]{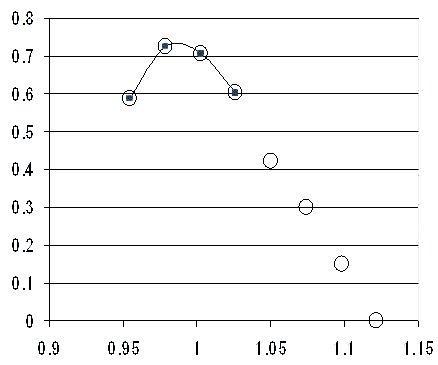}  
	  \label{fig2b}
}
\label{fig2}
\caption{The spectrum $f_\kappa(\alpha)$ for $\kappa = \kappa_c = 4.93g$ and frequency $\nu = 20$Hz}
\subfigure{
    \includegraphics[clip,height=2.0in,width=2.0in]{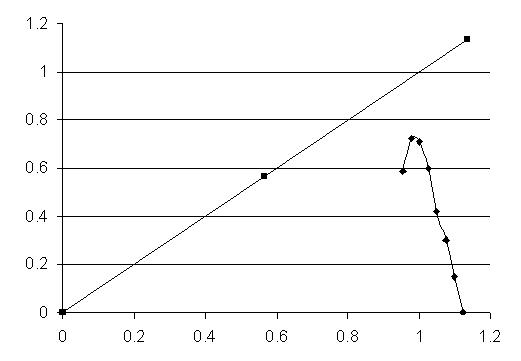}  
	  \label{k_493_100}
}
\label{k_493_100}
\caption{The spectrum $f_\kappa(\alpha)$, $\kappa = \kappa_c = 4.93g,$ $100$ boxes. }
\end{center}
\end{figure}
The segmented part of the graph implies a breakdown of the thermodynamical formalism, for the Lagrangian coordinate $q=f'(\alpha)$ collapses: if $q$ is constant, it cannot be the parametric coordinate of both $\alpha$ and $f(\alpha)$ shown in the figure. The segmented characteristic corresponding to this value $\kappa_c$ of $\kappa$ is stable under variations of $\mathbf{B}$: we observe (see Fig. 8) that, as before, the graph is not tangent to the bisectrix $y=x$. 
We increase $\mathbf{B}$ to $150$ (see Fig. 9), some irregularities appear in the curve; still the segmented part remains quite apparent.
\begin{figure}
\begin{center}
\subfigure{
    \includegraphics[clip,height=2.0in,width=2.0in]{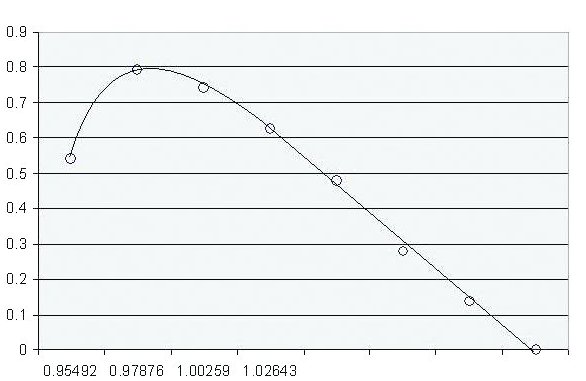}  
	  \label{k_493_150}
}
\label{k_493_150}
\caption{ The spectrum $f_\kappa(\alpha)$, $\kappa = \kappa_c = 4.93g$, $150 $ boxes. }
\subfigure{
    \includegraphics[clip,height=2.0in,width=2.0in]{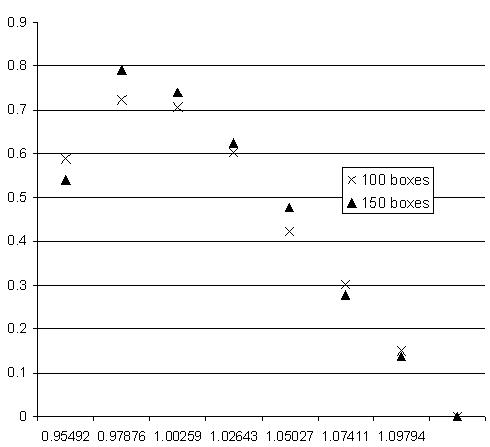}  
	\label{k_493_100_150}
}
\caption{ The spectra $f_\kappa(\alpha)$, for $\kappa = \kappa_c = 4.93g$, $100$ and $150 $ boxes. }
\end{center}
\end{figure}
Although the values of $\alpha_{min}$ and $\alpha_{max}$ for the two graphs of $f_{\kappa_c}(\alpha)$, for $\mathbf{B}=100$ and $\mathbf{B}=150$, are much the same, as seen in Fig. 10, Table 2 shows that for $\mathbf{B}=150$ the curve moves upwards, nearing line $y=x$. Moreover, $\alpha_M(\mathbf{B}=100)> \alpha_M(\mathbf{B}=150)$, which implies that the upper cusp of the curve moves towards the left (when $\mathbf{B}$ grows), again nearing $y=x$. As before, the \textquotedblleft distance" between the spectrum and line $y=x$ is only apparent, due to the low value of $\mathbf{B}$. For $\mathbf{B}=200$ we still have the segment, and the graph approaches even more the bisectrix,\dots\space but the curve begins to look shaky, losing its smoothness.
\newline\indent
The breakdown of the system, as source of information, goes beyond the collapse of the Lagrangian coordinate \textemdash i.e. $f'(\alpha)=constant$. The range of internal energies $\alpha$ allowed is $\Delta \alpha = \alpha_{max}- \alpha_{min} \cong 1.133-0.943=0.19$, whereas from Table 1, $\Delta\alpha$ is larger than $0.22$: the range of allowed internal energies is minimal for $\kappa=\kappa_c$, restricting further the information obtained from the spectrum. Also, $\alpha_M$, for which $f(\alpha)$ reaches its maximal value, is much nearer unity for $\kappa=\kappa_c$ than for $\kappa=4.08g$, as seen in Tables 1 and 2. The value $\alpha=1$ yields no information, since it means a \emph{uniform} distribution of probability vis-a-vis length, hence, a minimal variation in such distribution.
\begin{center}  
\begin{table}
  \begin{tabular*}{0.75\textwidth}%
     {@{\extracolsep{\fill}}cccr}
  --- & $\mathbf{B}=100$ & $\mathbf{B}=150$ \\
  \hline  
  $\alpha_{M} (f_{max})$  & 0.9788  & 0.9762   \\
  \hline  
  $f(\alpha_M)=f_{max}$  & 0.7235  & 0.7923   \\
  \end{tabular*} 
\caption{Maximal values for $f_{\kappa_c}(\alpha)$, $\mathbf{B}=100$ and $150$.}
\end{table} 
\end{center}
\section{Segmentation and the Henon map}
A multifractal spectrum with a possible segmented part was studied by Gunaratne and Procaccia (1987): the Henon map. In this system, anomalous periodic orbits with atypical behaviour contribute anomalous and isolated values of $\alpha$, and the end of the smooth part of the $f(\alpha)$ spectrum can be connected to some isolated $(\alpha,f(\alpha))$ point by a segment; $f'(\alpha)$ remaining continuous. The anomalous cycles contribute to isolated measures which, in the global partition sum from the thermodynamical formalism, yield such isolated $(\alpha,f(\alpha))$ that appear to suggest a segmented portion of the global spectrum. In the Faraday experiment we find that $f_{\kappa_c}(\alpha)$ is segmented because other measures get integrated in the global partition: for $\kappa>\kappa_c$ we find more than one measure clearly identified, contributing to the global multifractal algorithm.
\newline
\indent
In the Henon case, the segmented $f(\alpha)$ is conjectured to represent a possible phase transition \textemdash as previously suggested by Grassberger, Badii, and Politi (pers. comm. to Gunaratne and Procaccia). This segmented situation is deemed to be quite marginal, being extremely sensitive to minute variations of the Henon map's parameters. We interpret the segmented curve in Fig. 7 much in the same terms, for $f(\alpha)=f_{\kappa_c}(\alpha)$ is a marginal state extremely sensitive to minute variations of $\kappa$, since for $\kappa<\kappa_c$ there is no segment, and for $\kappa > \kappa_c$ there is no $\mu_k$, but two distinct measures (bi-multifractality, as seen below) which can be interpreted as a phase transition, extending the conjecture of Grassberger, Badii, and Politi.
\section{Post-crisis case $\kappa>\kappa_c$: Bi-multifractality}
Radons and Stoop (1996) studied the phenomenon of bi-multifractality obtained when two measures compete in the same support. They superposed two independent strictly self-similar multifractal measures supported in two Cantordust sets in the same segment. The two measures yield independent spectra $f_1(\alpha)$ and $f_2(\alpha)$, theoretically known, since they are self-similar. The references quoted in their work relate previous attempts at such superpositions. In Fig. 3b) in their paper, we observe the three spectra: the \emph{theoretically} known $f_1(\alpha)$ and $f_2(\alpha)$ in faded dotted lines and, in solid lines, the global generalised thermodynamical formalism that yields the empirical $f(\alpha)$, obtained by integrating the data from both measures: covering the whole segment that supports both measures with boxes and carrying out the multifractal algorithm. The resulting global integrated $f(\alpha)$ traces a bit of $(\alpha,f_1(\alpha))$ and a bit of $(\alpha,f_2(\alpha))$, with a substantial gap between the two broken, unconnected fragments: bi-multifractality. The graphs of $f_1(\alpha)$ and $f_2(\alpha)$ intersect at two points $P_l$ and $P_r$, \textquotedblleft $l$" and \textquotedblleft $r$" for \textquotedblleft left" and \textquotedblleft right". One of the stopping points of the fragmented global $f(\alpha)$ is $P_r$, where $f'_1$ and $f'_2$ are finite and have different signs. This is interpreted by Radons and Stoop as a change of phase, second-order transition. 
\begin{figure}
\begin{center}
\subfigure[For $\nu = 20$ Hz, $\kappa = 6.57g > \kappa_c$ yields a second embryonic spectrum]{
    \includegraphics[clip,height=1.5in,width=1.5in]{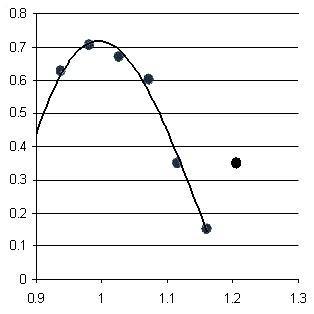} 
	  \label{fig3a}
}
\subfigure[We enhance $\kappa$ and $\nu$: $\kappa = 8.95$g and $\nu=30$Hz yield two fragmented spectra]{
    \includegraphics[clip,height=1.5in,width=1.5in]{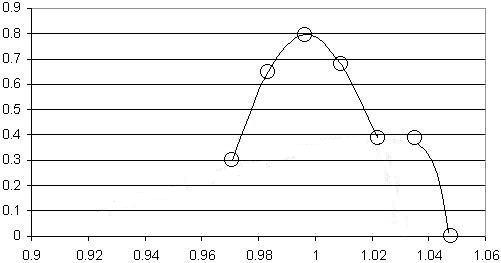}  
	  \label{fig3b}
}
\subfigure[Large $\kappa$ and $\nu$ values for the same liquid but different viscosity yield again two fragmented spectra]{
    \includegraphics[clip,height=1.5in,width=1.5in]{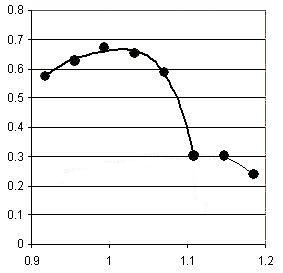}  
	  \label{fig3c}
}
\label{fig3}
\caption{Fragmented spectra (bi-multifractality) for high values of $\kappa$}
\end{center}
\end{figure}
Fig. 11 shows global integrated spectra taken from the Faraday experiment, for $\kappa>\kappa_c$ in each case. Fig. 11 a) shows an embryonic $f_2(\alpha)$. Figs. 11 b) and c) show unconnected fragments of two multifractal spectra, separated by a gap: bi-multifractality. Neither of the two fragments of spectrum shows the symmetries corresponding to self-similar multifractal characteristics. But the geometrical features common to the pairs of fragments in Figs. 11 a), b) and c), suggest a sketching of a bi-multifractal configuration \textemdash see Fig. 12\textemdash\space in accordance with the second-order transition of the Radons and Stoop model. 
\begin{figure}
\begin{center}
\subfigure{
    \includegraphics[clip,height=1.5in,width=1.5in]{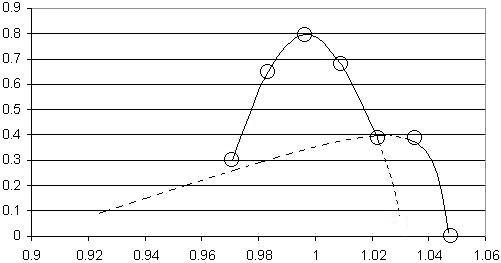}  
	  \label{fig3bp}
}
\subfigure{
    \includegraphics[clip,height=1.5in,width=1.5in]{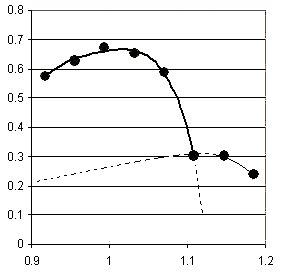}  
	  \label{fig3cp}
}
\label{fig3p}
\caption{Bi-multifractal configurations \textemdash Figs. 11(b) and 11(c)\textemdash\space in accordance with the Radons and Stoop model}
\end{center}
\end{figure}
The shape of the fragmented spectra $f_\kappa, \kappa>\kappa_c$ in Figs. 11 b) and c), corresponding to different cases, suggests a \emph{common road} post-crisis-through-bi-multifractality for this Faraday experiment. 
\newline
\indent
In the Henon case, the fractal attractor $F_\kappa$ disappears for $\kappa \geq \kappa_c$; analogously, the post-crisis type of situation for $\kappa \geq \kappa_c$ in the Faraday experiment shows that the multifractal measure $\mu_\kappa$ disappears \textemdash for the set of thermodynamical laws ceases to be valid for the broken $f_{\kappa}(\alpha)$. This disappearance of $\mu_\kappa$ is explained by the presence of two (or more?) conflicting measures.
\newline
\indent
There is a continuity among the chaos-road-to-crisis, crisis, and post-crisis-bi-multifractality cases: the first one, $\kappa
<\kappa_c$, shows a \textquotedblleft normal" $\mu_\kappa$ through a \textquotedblleft normal" $f(\alpha)$; the second case, $\kappa=\kappa_c$, shows anomalous measures, yielding a not normal spectrum through the appearance of a segment, which is interpreted as a change of phase. The third case shows, explicitly and clearly, those conflicting measures: the spectrum is not \textquotedblleft anomalous" any more, but broken down, the phase transition now quite apparent. 
\section{Qualitative observations}
Past the crisis point, we obtain, in the three cases of Fig. 11, a bi-multifractal situation. The three cases seem to share some qualitative \textemdash if not quantitative\textemdash\space properties. The case of Fig. 11 a) is the weakest, since the second multifractal reduces itself to a single point. Yet, it is a point that cannot be ignored: on the one hand, it is not on the real $\alpha-$axis, which means that it is not on a freak isolated box with just one lonely element in the Cantorset, the other $9,999$ being together elsewhere. On the other hand, it is far enough away from the typical and usual $(\alpha,f(\alpha))$ at its left, as to make it impossible to consider this point as part of that convex curve. The one-point spectrum is exactly at the height of one of the points in the typical graph at the left, a feature also present in the other two graphs \textemdash Figs. 11 b) and c). With higher values of the parameters, the two broken spectra (Fig. 11 b)) appear more clearly, yielding bi-multifractality. Fig. 11 c) presents the same type of bi-multifractality (from a qualitative point of view), for the same liquid at a higher viscosity. A word on the significance of this fact: as an example, let us consider the Farey tree (Piacquadio and Rosen 1995, see also Piacquadio and Cesaratto 2001). The tree is a partition of unit segment $[0,1]=[\frac{0}{1},\frac{1}{1}]$, obtained by interpolating rational number $\frac{p+p'}{q+q'}$ between $\frac{p}{q}$ and $\frac{p'}{q'}$. Successive interpolations give rise to all rational numbers, starting from $\frac{1}{2}=\frac{0+1}{1+1}$ interpolated between $\frac{0}{1}$ and $\frac{1}{1}$. A certain optoelectronic experiment (Essevaz-Roulet \emph{et al.} 2000) shows a camera that films a screen to which it is connected, filming its own image, feeding back the image to the screen. When the camera turns around its optical axis, a pattern of $p$ light spots appear on the screen, after $q$ turns. Integers $p$ and $q$ follow the hierarchy of the Farey tree\dots\space But, when measuring $p$'s and $q$'s, the precision of such measurements has to increase with $q$, so the Farey tree has a \emph{natural} bound for the $q$'s in each refinement of the partition: all rational numbers $\frac{p}{q}$ \emph{will} appear, but they do so slower than in the theoretical interpolation, for they do it, perforce, according to the size of $q$. All other things remain equal: the interpolation laws in the unit segment, the equidistribution of the probability measure\dots\space Yet, when processing the theoretical partition (Farey-Brocot tree) as a multifractal spectrum $f_{F-B}(\alpha)$, and the practical one $f_F(\alpha)$, perforce a slower production of rational numbers \textemdash due to inescapable experimental reasons\textemdash\space we find that $f_{F-B}(\alpha)$ and $f_F(\alpha)$ cannot be more different. The theoretical $f_{F-B}(\alpha)$ is strictly increasing and the range of the $\alpha$'s is $[\alpha_{min},\infty)$: an infinite segment; whereas $f_F(\alpha)$ is strictly decreasing, the $\alpha$ range is $[1,2]$. Hence, the same interpolation rules, the same measure distribution rules, applied to the same fractal support,\dots\space with an inescapable slow-down in the velocity of interpolation, yield totally unrelated multifractal spectra. Our point being that the algorithm yielding a multifractal spectrum is extremely sensitive to (apparently innocuous) variations in its measure. Hence, we can conjecture that, when two spectra have similar qualitative characteristics, as is the case illustrated in Figs. 11 b) and c), the underlying phenomena studied with multifractal tools have to be strongly connected. 
\section{Discussion and Conclusions}

For different values of $\kappa$, the processed data of the corresponding signals \textemdash according to the frame described above\textemdash\space agree with direct observations of the experiment: for $\kappa \nearrow \kappa_c$ (chaos-road-to-crisis) we observe the cusps maintaining the same locus on the surface and the same height.
For $\kappa > \kappa_c$ (bi-multifractality) the cusps are sparse and do not preserve their location on the surface. They group in clusters that \textquoteleft wander around\textquoteright \ on the liquid surface. When we arrive at 30 Hz and $\kappa >> \kappa_c$ the cusps are sporadic and appear to be random. Beyond bi-multifractality we observe the ejection of droplets (spray). For $\mathbf{S} > 10^4$ the liquid starts consistently losing cusps and the system tends to change its structure, as shown in Fig. 13.
\newline
\indent
The laboratory experiment perforce limits the size of the data set available for analysis. Despite this limitation, multifractal data processing permits the identification of distinct states (chaos, crisis, \dots). These (laboratory) results are in accordance with pre-existing theoretical models of such states. Especially noteworthy is that the different states can be visually observed in the course of the experiment. 
\begin{figure}
\begin{center}
    \includegraphics[clip,height=2.05in,width=5.0in]{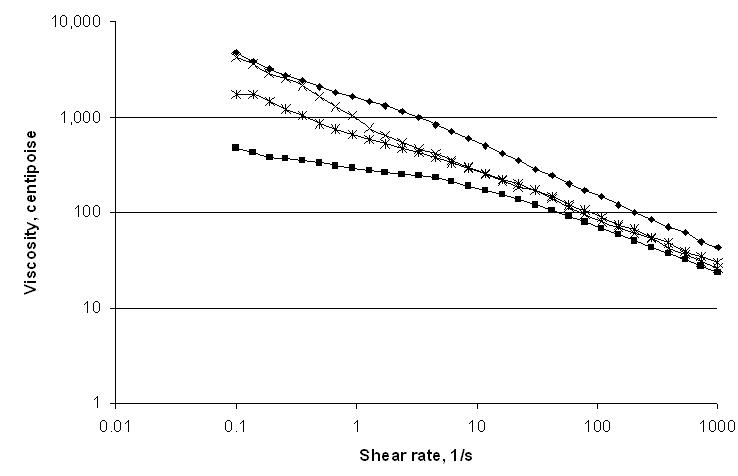} 
	  \label{h}
\caption{Different viscosity/shear-rate curves imply that the liquid has become non-homogeneous for $\mathbf{S} > 10^4$.}
\end{center}
\end{figure}
\section{Acknowledgements}
We are indebted to Carlos Lizarralde for his technical assistance in measurements and data processing, for helpful suggestions, and for encouragements too numerous to mention.

\end{document}